\documentclass[aps,prl,reprint,groupedaddress]{revtex4-1}
\usepackage{amsmath}
\usepackage{array}
\usepackage{multirow}
\usepackage{enumitem}

\input{tcilatex}

\begin{document}

\title{Classification of general $n$-qubit states under stochastic local 
operations and classical communication in terms of the
rank of coefficient matrix}
\author{Xiangrong Li$^1$, Dafa Li$^{2,3}$}
\address{$^1$Department of Mathematics, University of California Irvine,
Irvine, California 92697, USA\\
$^2$Department of Mathematical Sciences, Tsinghua University,
Beijing, 100084, China\\
$^3$Center for Quantum Information Science and Technology, Tsinghua National
Laboratory for Information \\
Science and Technology (TNList), Beijing, 100084, China}

\begin{abstract}
We solve the entanglement classification under stochastic local operations
and classical communication (SLOCC) for general $n$-qubit states. For two
arbitrary pure $n$-qubit states connected via local operations, we establish
an equation between the two coefficient matrices associated with the
states. The rank of the coefficient matrix is preserved under SLOCC
and gives rise to a simple way of partitioning all the pure
states of $n$ qubits into different families of entanglement classes,
as exemplified here.
When applied to the symmetric states, this
approach reveals that all the Dicke states $|\ell,n\rangle$ with $%
\ell=1,\dots, [n/2]$ are inequivalent under SLOCC.
\end{abstract}

\maketitle


\affiliation{$^1$ Department of Mathematics, University of
California, Irvine, CA 92697-3875, USA \\
$^2$ Department of mathematical sciences, Tsinghua University,
Beijing 100084 CHINA}



\section{I. Introduction}

Entanglement, a key feature that distinguishes quantum information from
classical information, has applications in cryptography, teleportation and
quantum computation \cite{Nielsen}. While the bipartite entanglement is well
understood, the task of classifying multipartite entanglement beyond two
qubits becomes increasingly difficult. To classify entangled states, some
equivalence relation has to be introduced. Of particular importance is the
equivalence under stochastic local operations and classical communication
(SLOCC) since two states belonging to the same equivalence class can perform
the same tasks of quantum information theory.

For three qubits, in terms of the local ranks of the reduced density
matrices,
it has been shown that there are six inequivalent SLOCC
classes \cite{Dur}. For four or more qubits, there exists an infinite number of
inequivalent SLOCC classes. It is highly desirable to partition the infinite
SLOCC classes into a finite number of families such that states belonging to
the same family possess similar properties, according to some criteria for
determining which family a given state belongs to. Considerable efforts have
been undertaken over the last decade for the SLOCC entanglement
classification of four-qubit states resulting in a finite number of
families \cite{Moor2,Chterental,Lamata07,Borsten} or classes \cite%
{Cao,LDF07b,LDFQIC,Buniy,Viehmann,Zha}. For more than four qubits, a few
attempts have been made for SLOCC classification for subsets of the
general $%
n$-qubit states such as the Greenberger-Horne-Zeilinger 
(GHZ)-type, W-type, and GHZ-W-type $n$-qubit states
\cite{Chen}, symmetric $n$-qubit states \cite{LDFEPL,Bastin,Aulbach}, even $n
$-qubit states \cite{LDFJPA,LDFJPA12}, and odd $n$-qubit states \cite{LDFQIC11}%
. Despite these efforts, a SLOCC classification for general $n$-qubit states
is still beyond reach.

Our aim is to solve the problem of SLOCC classification for all multipartite
pure states in the general $n$-qubit case. To this end, we demonstrate that the
rank of the coefficient matrix of a pure $n$-qubit state is invariant under
SLOCC. SLOCC invariants for subsets of $n$-qubit states
have been the subject of several recent studies
\cite{LDF07a,LDFJPA,LDFJPA12,LDFQIC11,Ribeiro}.
In \cite{LDFJPA,LDFJPA12}, the invariant element is the
determinant of coefficient matrices of even $n$ qubits.
In \cite{LDFQIC11}, the invariant element is the rank of square matrices
of size two constructed using three functions defined on the space of odd $n$
qubits.

We construct the coefficient matrices for general $n$-qubit
states by arranging the coefficients in lexicographical order.
For two states connected via local operations, their coefficient
matrices are related through an equation.
In the case where the local operations are invertible,
the two states are said to be SLOCC equivalent and the two coefficient
matrices
have the same rank; i.e., the rank is preserved under SLOCC.
The rank gives rise to a simple way of partitioning all the $n$-qubit states
into different SLOCC families.
For $n$-qubit symmetric Dicke states $|\ell,n\rangle$ with $\ell$
($\ell=1,\dots, [n/2]$) excitations,
we show that the rank of the coefficient matrix of $|\ell,n\rangle$
is equal to $\ell+1$ and, therefore, all these states are inequivalent
under SLOCC. Finally, composing the rank and permutations of qubits
allows us to define subfamilies by cutting each family in pieces.

\section{II. SLOCC matrix equation and the invariance of the rank}

We write the state $|\psi ^{\prime }\rangle $ of $n$ qubits as $|\psi
^{\prime }\rangle =\sum_{i=0}^{2^{n}-1}a_{i}|i\rangle $, where $|i\rangle $
are basis states and $a_{i}$ are coefficients. We associate to an $n$-qubit
state $|\psi ^{\prime }\rangle $ a $2^{[n/2]}\times 2^{[(n+1)/2]}$
coefficient matrix $M(|\psi ^{\prime }\rangle )$ whose entries are the
coefficients $a_{0},a_{1},\dots ,a_{2^{n}-1}$ arranged in ascending
lexicographical order. To illustrate, we list below $M(|\psi ^{\prime
}\rangle )$ for $n=3$:
\begin{equation}
M(|\psi ^{\prime }\rangle )=\left(
\begin{array}{cccc}
a_{0} & a_{1} & a_{2} & a_{3} \\
a_{4} & a_{5} & a_{6} & a_{7}%
\end{array}%
\right),  \label{M3}
\end{equation}
and for $n=4$:
\begin{equation}
M(|\psi ^{\prime }\rangle )=\left(
\begin{array}{cccc}
a_{0} & a_{1} & a_{2} & a_{3} \\
a_{4} & a_{5} & a_{6} & a_{7} \\
a_{8} & a_{9} & a_{10} & a_{11} \\
a_{12} & a_{13} & a_{14} & a_{15}%
\end{array}%
\right).  \label{M4}
\end{equation}

We refer to the rank of the coefficient matrix $M(|\psi ^{\prime }\rangle )$
as the rank of the state $|\psi ^{\prime }\rangle $, denoted as rank$(|\psi
^{\prime }\rangle )$. We exemplify with the $n$-qubit $|\mbox{GHZ}\rangle $ 
state $\frac{1}{\sqrt{2}}(|0\rangle ^{\otimes n}+|1\rangle ^{\otimes n})$, 
and we find rank$(|\mbox{GHZ}\rangle )=2$. 
It is clear that the rank of any $n$-qubit
state ranges over the values 1, 2, $\dots $, $2^{[n/2]}$.

\textsl{Theorem.} Let $|\psi \rangle $ be another state of $n$ qubits with $%
|\psi \rangle =\sum_{i=0}^{2^{n}-1}b_{i}|i\rangle $ and $M(|\psi \rangle )$
be the corresponding coefficient matrix constructed in the same manner as
was done for $M(|\psi ^{\prime }\rangle )$. If the states $|\psi \rangle $
and $|\psi ^{\prime }\rangle $ are related by
\begin{equation}
|\psi ^{\prime }\rangle =\mathcal{A}_{1}\otimes \mathcal{A}_{2}\otimes
\mbox{\dots} \otimes \mathcal{A}_{n}|\psi \rangle ,  \label{g-relation}
\end{equation}%
where the local operators $\mathcal{A}_{1}$, $\mathcal{A}_{2},\dots $,
and $%
\mathcal{A}_{n}$ are not necessarily invertible, then the following
matrix equation holds for general $n$ qubits:
\begin{equation}
M(|\psi ^{\prime }\rangle )=(\mathcal{A}_{1}\otimes \mbox{\dots} \otimes
\mathcal{A%
}_{[n/2]})M(|\psi \rangle )(\mathcal{A}_{[n/2]+1}\otimes \mbox{\dots} \otimes
\mathcal{A}_{n})^{T}.  \label{matrix-1}
\end{equation}
Equation (\ref{matrix-1}) holds particularly true for two SLOCC equivalent states
$|\psi \rangle $ and $|\psi ^{\prime }\rangle $ which satisfy Eq. (\ref%
{g-relation}) along with the local operators $\mathcal{A}_{1}$, $\mathcal{A}%
_{2},\dots $, and $\mathcal{A}_{n}$ being invertible \cite{Dur}. It follows
from Eq. (\ref{matrix-1}) that two SLOCC equivalent states have the same
rank, in other words, the rank is invariant under SLOCC, thereby revealing
that the rank is an inherent physical property. Therefore, if two states
differ in their ranks, then they belong necessarily to different SLOCC
equivalent classes. It should be noted that the converse does not hold;
i.e., two states with the same rank are not necessarily equivalent.

We have the following two simple results:
(i) The rank of a full separable state is always 1.
(ii) The rank of a genuinely entangled state is always greater than 1.

\textsl{Remark 1.} Taking the determinants of both sides of Eq. (\ref%
{matrix-1}) for even $n$ yields \cite{LDFJPA,LDFJPA12}:
\begin{equation}
\det M(|\psi^{\prime }\rangle)=\det M(|\psi\rangle)[\det (\mathcal{A}%
_1)\dots \det(\mathcal{A}_n)]^{2^{(n-2)/2}}.  \label{matrix-2}
\end{equation}
It follows from Eq. (\ref{matrix-2}) that if one of $\det M(|\psi^{\prime
}\rangle)$ and $\det M(|\psi\rangle)$ vanishes while the other does not,
then the state $|\psi^{\prime }\rangle$ is not equivalent to $|\psi\rangle$
under SLOCC.
In view of the fact that the determinant of a matrix is nonvanishing if and
only if it has full rank, the SLOCC invariance of the rank is stronger than
the invariance of the determinant.

\section{III. SLOCC classification in terms of the rank}

We define the family $\mathcal{F}_{n,r}$ to be the set of all $n$-qubit
states with the same rank $r$. In the sequel, we will omit the subscript $n$
and simply write $\mathcal{F}_r$, whenever the number of qubits is clear
from the context. Thus, there exist $2^{[n/2]}$ different SLOCC families for
any $n$ qubits. Clearly, if two states are SLOCC equivalent then they belong
to the same family. However, the converse does not hold: two states
belonging to the same family may be inequivalent under SLOCC. It is further
noted that when $n\ge 4$, at least one family contains an infinite number of 
SLOCC classes.

For any $n$ qubits, the following hold:
(i) Family $\mathcal{F}_{1}$ contains all the full separable states.
(ii) Family $\mathcal{F}_{1}$ contains no genuine entangled states.
(iii) Family $\mathcal{F}_{1}$ contains finite SLOCC classes.
(iv) Family $\mathcal{F}_{2}$ contains the $n$-qubit $|\mbox{GHZ}\rangle $ state.
(v) Family $\mathcal{F}_{2+r}$ contains the following state:
\begin{equation}
\frac{1}{\sqrt{r+2}}(|0\rangle -|2^{n}-1\rangle
+\sum_{k=1}^{r}|k(2^{[(n+1)/2]}+1)\rangle),
\end{equation}
where $1\leq r\leq 2^{[n/2]}-2$.

Now we turn to the $n$-qubit symmetric Dicke states $|\ell,n\rangle$ with $%
\ell$ excitations \cite{Stockton}:
\begin{equation}
|\ell,n\rangle =\left( _{\ell}^{n}\right) ^{-1/2}\sum\limits_k
P_{k}|1_{1},1_{2},\dots, 1_{\ell},0_{\ell+1},\dots, 0_{n}\rangle,
\label{Dicke}
\end{equation}
where $\ell$ ranges from 1 to $n-1$ and $\{P_{k}\}$ is the set of all
distinct permutations of the spins. These states have been featured in
theoretical studies \cite{Toth,Huber} and implemented experimentally \cite%
{Prevedel,Wieczorek}. The Dicke state $|1,n\rangle$ is just the $n$-qubit $%
|W\rangle$ state and $|\ell,n\rangle $ is equivalent to $|n-\ell ,n\rangle $
under SLOCC.

As has been previously noted, all symmetric Dicke states $|\ell ,n\rangle $
with $\ell =1,\dots,[n/2]$ are SLOCC inequivalent \cite{LDFEPL,Bastin}.
These states, as demonstrated below, can also be distinguished by the rank
of their coefficient matrices which depends only on the number of excitations
and is independent of the number of qubits.
Since rank$(|\ell ,n\rangle )=$rank$(|n-\ell,n\rangle )$,
we only need to compute rank$(|\ell ,n\rangle )$ with $1\leq
\ell \leq \lbrack n/2]$. This can be done as follows. We first construct the
coefficient matrix $M(|\psi^{\prime }\rangle )$ of state $|\psi^{\prime
}\rangle$ in the same manner as discussed above. We may write $|\psi^{\prime
}\rangle $ in terms of an orthogonal basis as $|\psi^{\prime }\rangle =\sum
a_{i_{1}i_{2}\dots i_{n}}|i_{1}i_{2}\dots i_{n}\rangle$, where $%
i_{1}i_{2}\dots i_{n}$ is the $n$-bit binary form of the index $i$.
Inspection of the structure of the matrix $M(|\psi ^{\prime }\rangle )$
reveals that the coefficient $a_{i_{1}\dots i_{[n/2]}i_{[n/2]+1}\dots
i_{n}}$ is the entry in the $(i_{1}\dots i_{[n/2]})$th row and $%
(i_{[n/2]+1}\dots i_{n})$th column of the matrix. Here, the $n$ bits are
split into two halves, referred to as the row bits and column bits,
respectively: bits $1$ to ${[n/2]}$ are used to specify the row number, and
bits $[n/2]+1$ to $n$ are used to specify the column number. In view of
Eq. (%
\ref{Dicke}), the nonzero entries of the coefficient matrix $%
M(|\ell,n\rangle )$ are those whose $n$-bit binary forms have $\ell $ bits
equal to $1$ and the rest of the bits equal to $0$. We further observe that the
rows of $M(|\ell ,n\rangle )$ with no more than $\ell $ row bits equal to 1
are nonzero rows, while the remaining rows are identically zero. Consider
the rows with $j$ ($0\leq j\leq \ell $) row bits equal to 1. Clearly, there
are $\binom{n/2}{j}$ such rows of $M(|\ell ,n\rangle )$ that are identical.
Letting $j$ vary from 0 to $\ell $ gives a total of $\ell +1$ different
rows. It can be verified that these $\ell +1$ rows are independent. This
yields $rank(|\ell ,n\rangle )=\ell +1$.

Accordingly, for any $n$ qubits, all the Dicke states $|\ell ,n\rangle $
with $\ell =1$, $\dots $, $[n/2]$, are inequivalent under SLOCC, since they
differ in their ranks. Further, it can readily be seen that the Dicke
state $%
|\ell ,n\rangle $ with $\ell =1$, $\dots $, $[n/2]$\ belongs to the
family $%
\mathcal{F}_{\ell +1}$. This gives rise to a complete SLOCC classification
of all the symmetric Dicke states.

\textsl{Remark 2.} It follows from the discussion above that both
the states $|W\rangle $ and $|\mbox{GHZ}\rangle$ have the same rank, thereby
revealing that the two states have a similar algebraic structure.
It is further noted that both the states $|W\rangle $ and $|\mbox{GHZ}\rangle$ 
admit a similar Frobenius algebra structure \cite{Coecke}.

\section{IV. Ranks of coefficient matrices under permutations of qubits}

In \cite{LDFJPA,LDFJPA12}, we presented a systematic way to find all the
possible coefficient matrices for even $n$-qubit states such that the
determinants of these coefficient matrices are invariant under SLOCC. Here
we extend this construction to general $n$ qubits. Observe that to write a $%
2^{[n/2]}\times 2^{[(n+1)/2]}$ matrix into binary index form, we need $[n/2]$
row bits and $[(n+1)/2]$ column bits. In the binary form of the coefficient
matrix given in Eqs. (\ref{M3}) and (\ref{M4}), bits $1$ to ${[n/2]}$ are the 
row bits, and bits $[n/2]+1$ to $n$ are the column bits. Alternatively, we may
choose any $[n/2]$ bits as the row bits and the remaining $[(n+1)/2]$ bits
as the column bits. This amounts to
$(\frac{1}{2})^{n+1\bmod{2}}\binom{n}{[n/2]}$
different coefficient matrices, ignoring those that end up exchanging
rows or columns.
The factor of $1/2$ for even $n$ arises because exchanging the 
row and column bits is equivalent to transposing the matrix.
It turns out that the ranks of these coefficient matrices are all
invariant under SLOCC.
To see this, we will resort to
permutations of qubits. Let $\sigma $ be a permutation of qubits given by
\cite{LDFJPA12}
\begin{equation}
\sigma =(q_{1},t_{1})(q_{2},t_{2})\dots (q_{k},t_{k}),  \label{eq_sigma}
\end{equation}%
where $(q_{i},t_{i})$ is the transposition of a pair of qubits $q_{i}$ and 
$t_{i}$ with $q_{i}$ being a row bit and $t_{i}$ a column bit. Exhausting all
possible values of $q_{1},\dots ,q_{k}$ and $t_{1},\dots ,t_{k}$ such that
$1\leq q_{1}<q_{2}<\mbox{\dots} <q_{k}<[(n+1)/2]$, $[n/2]<t_{1}<t_{2}<\mbox{\dots}<t_{k}\leq n$, and letting $k$ vary from 0 to $[(n-1)/2]$ 
(we define $\sigma =I$ for $k=0$), 
yields $(\frac{1}{2})^{n+1\bmod{2}}\binom{n}{[n/2]}$ different permutations of
qubits. Let $M^{\sigma }(|\psi ^{\prime }\rangle )$ denote the coefficient
matrix of the state $|\psi ^{\prime }\rangle $ under permutation $\sigma $,
and let rank$^{\sigma }(|\psi ^{\prime }\rangle )$ denote its rank. We may
refer to rank$^{\sigma }(|\psi ^{\prime }\rangle )$ as the rank of the state
$|\psi ^{\prime }\rangle $ under permutation $\sigma $. Simply taking the
permutation $\sigma $ to both sides of Eq. (\ref{matrix-1}) yields the
following SLOCC matrix equation:
\begin{eqnarray}
M^{\sigma}(|\psi^{\prime}\rangle)
&=&(\mathcal{A}_{\sigma(1)}\otimes \mbox{\dots} \otimes
\mathcal{A}_{\sigma([n/2])})M^{\sigma}(|\psi\rangle)\notag\\
& & \quad \quad  (\mathcal{A}_{\sigma([n/2]+1)}\otimes \mbox{\dots} \otimes
\mathcal{A}_{\sigma(n)})^{T}.  \label{ext-1}
\end{eqnarray}

It follows immediately from Eq. (\ref{ext-1}) that two SLOCC equivalent
states have the same rank with respect to the permutation $\sigma $. That
is, the rank with respect to the permutation $\sigma $ is invariant under
SLOCC. Conversely, if two states differ in their ranks with respect to the
permutation $\sigma$, then they belong necessarily to different SLOCC
classes.
We define the family $\mathcal{F}_{r}^{\sigma}$ to be the set of all
$n$-qubit
states with the same rank $r$ with respect to the permutation $\sigma$,
where $r$ ranges from 1 to $2^{[n/2]}$ and we have omitted a subscript $n$.
Suppose $\sigma_1,\sigma_2,\dots,\sigma_m$ with 
$m\leq (\frac{1}{2})^{n+1\bmod{2}}\binom{n}{[n/2]}$
is a sequence of permutations of the form given in Eq. (\ref{eq_sigma}).
In terms of the rank of $M^{\sigma_1}$, the $n$-qubit states are divided into
$2^{[n/2]}$ families: $\mathcal{F}_{r_1}^{\sigma_1}$.
Then, in terms of the rank of $M^{\sigma_2}$,
each family $\mathcal{F}_{r_1}^{\sigma_1}$
can be further divided into $2^{[n/2]}$ subfamilies:
$\mathcal{F}_{r_1,r_2}^{\sigma_1\sigma_2}=\mathcal{F}_{r_1}^{\sigma_1}
\cap \mathcal{F}_{r_2}^{\sigma_2}$.
Here, each subfamily $\mathcal{F}_{r_1,r_2}^{\sigma_1\sigma_2}$ is the
intersection of the families $\mathcal{F}_{r_1}^{\sigma_1}$ and
$\mathcal{F}_{r_2}^{\sigma_2}$.
Assume that in terms of the ranks of
$M^{\sigma _{1}},M^{\sigma _{2}},\dots ,M^{\sigma _{m-1}}$, the $n$-qubit
states are divided into $2^{(m-1)[n/2]}$ families: $\mathcal{F}%
_{r_{1},r_{2},\dots ,r_{m-1}}^{\sigma _{1}\sigma _{2}\dots \sigma _{m-1}}=%
\mathcal{F}_{r_{1}}^{\sigma _{1}}\cap \mbox{\dots} \cap \mathcal{F}
_{r_{m-1}}^{\sigma _{m-1}}$. Then, in terms of the rank of $M^{\sigma _{m}}$%
, each family $\mathcal{F}_{r_{1},r_{2},\dots ,r_{m-1}}^{\sigma _{1}\sigma
_{2}\dots \sigma _{m-1}}$ can be further divided into $2^{[n/2]}$
subfamilies:
$\mathcal{F}_{r_{1},r_{2},\dots ,r_{m}}^{\sigma _{1}\sigma _{2}\dots
\sigma _{m}}=\mathcal{F}_{r_{1},r_{2},\dots ,r_{m-1}}^{\sigma _{1}\sigma
_{2}\dots \sigma _{m-1}}\cap \mathcal{F}_{r_{m}}^{\sigma _{m}}=\mathcal{F}%
_{r_{1}}^{\sigma _{1}}\cap \mbox{\dots} \cap \mathcal{F}_{r_{m}}^{\sigma _{m}}$.
This gives a total of $2^{m[n/2]}$ different SLOCC families.

We exemplify with the family $L_{a_{2}b_{2}}=a(|0000\rangle+|1111%
\rangle)+b(|0101\rangle+|1010\rangle) +|0011\rangle+|0110\rangle$ for four
qubits presented by Verstraete \emph{et al.} \cite{Moor2}. As shown in Table
\ref{table_verstraete}, the family $L_{a_{2}b_{2}}$ is further divided
into four
subfamilies (all other subfamilies are empty) with respect to permutations
$\sigma _{1}=I$ and $\sigma_{2}=(1,4)$:
$\mathcal{F}_{2,1}^{\sigma _{1}\sigma _{2}}$, $\mathcal{F}%
_{3,3}^{\sigma _{1}\sigma _{2}}$, and $\mathcal{F}_{4,2}^{\sigma _{1}\sigma
_{2}}$ contain only a single SLOCC class, while $\mathcal{F}_{4,3}^{\sigma
_{1}\sigma _{2}}$ contains an infinite number of SLOCC classes.
In a similar fashion, we can further divide other families presented by
Verstraete \emph{et al.} \cite{Moor2} into subfamilies.

\begin{table}[tbph]
\caption{SLOCC classification of $L_{a_{2}b_{2}}$}
\label{table_verstraete}%
\begin{ruledtabular}
\begin{tabular}{cccc}
$\mathcal{F}_{1}^{\sigma _{1}}$ & $\mathcal{F}_{2}^{\sigma _{1}}$ &
$\mathcal{F}_{3}^{\sigma _{1}}$ & $\mathcal{F}_{4}^{\sigma _{1}}$ \\
$\emptyset $ & $a=b=0$ & $ab=0$ \& $a\neq b$ & $ab\neq 0$ \\
$\mathcal{F}_{1}^{\sigma _{2}}$ & $\mathcal{F}_{2}^{\sigma _{2}}$ &
$\mathcal{F}_{3}^{\sigma _{2}}$ & $\mathcal{F}_{4}^{\sigma _{2}}$ \\
$a=b=0$ & $a=\pm b$ \& $a\neq 0$ & $a\neq \pm b$ & $\emptyset $ \\
$\mathcal{F}_{2,1}^{\sigma _{1}\sigma _{2}}$ &
$\mathcal{F}_{3,3}^{\sigma_{1}\sigma _{2}}$ &
$\mathcal{F}_{4,2}^{\sigma _{1}\sigma _{2}}$ &
$\mathcal{F}_{4,3}^{\sigma _{1}\sigma _{2}}$ \\
$a=b=0$ & $ab=0$ \& $a\neq b$ & $a=\pm b$ \& $a\neq 0$
& $ab\neq 0$ \& $a\neq \pm b$
\end{tabular}
\end{ruledtabular}
\end{table}

Consider the family span$\{0_k\Psi,0_k\Psi\}=
|0000\rangle+|1100\rangle+\alpha|0011\rangle+\beta|1111\rangle$ for four
qubits presented by Lamata \emph{et al.} \cite{Lamata07}. As shown in Table
\ref{table_lamata}, the family $\{0_k\Psi,0_k\Psi\}$ is further divided into
four subfamilies (all other subfamilies are empty) with respect to
permutations $\sigma _{1}=I$ and $\sigma_{2}=(1,4)$:
$\mathcal{F}_{1,2}^{\sigma _{1}\sigma _{2}}$,
$\mathcal{F}_{1,4}^{\sigma _{1}\sigma _{2}}$, and
$\mathcal{F}_{2,3}^{\sigma _{1}\sigma _{2}}$
contain only a single SLOCC class, while
$\mathcal{F}_{2,4}^{\sigma _{1}\sigma _{2}}$ contains an infinite number of
SLOCC classes.
In a similar way, other families presented by Lamata \emph{et al.}
\cite{Lamata07}
can also be further divided into subfamilies.

\begin{table}[tbph]
\caption{SLOCC classification of span$\{0_k\Psi,0_k\Psi\}$}
\label{table_lamata}%
\begin{ruledtabular}
\begin{tabular}{cccc}
$\mathcal{F}_{1}^{\sigma _{1}}$ & $\mathcal{F}_{2}^{\sigma _{1}}$ &
$\mathcal{F}_{3}^{\sigma _{1}}$ & $\mathcal{F}_{4}^{\sigma _{1}}$ \\
$\alpha=\beta$ & $\alpha\neq \beta$ & $\emptyset $ & $\emptyset $ \\
$\mathcal{F}_{1}^{\sigma _{2}}$ & $\mathcal{F}_{2}^{\sigma _{2}}$ &
$\mathcal{F}_{3}^{\sigma _{2}}$ & $\mathcal{F}_{4}^{\sigma _{2}}$ \\
$\emptyset $ & $\alpha=\beta=0$ &
$\alpha\beta=0$ \& $\alpha\neq \beta$ & $\alpha\beta\neq 0$ \\
$\mathcal{F}_{1,2}^{\sigma _{1}\sigma _{2}}$ &
$\mathcal{F}_{1,4}^{\sigma_{1}\sigma _{2}}$ &
$\mathcal{F}_{2,3}^{\sigma _{1}\sigma _{2}}$ &
$\mathcal{F}_{2,4}^{\sigma _{1}\sigma _{2}}$ \\
$\alpha=\beta=0$ & $\alpha=\beta\neq 0$ &
$\alpha\beta=0$ \& $\alpha\neq \beta $ &
$\alpha\beta\neq 0$ \& $\alpha\neq \beta$
\end{tabular}
\end{ruledtabular}
\end{table}

By using the filters, it has been shown that four
five-qubit states $|\Psi _{2}\rangle$, $|\Psi _{4}\rangle$,
$|\Psi _{5}\rangle$, and $|\Psi _{6}\rangle$
are in different orbits \cite{Osterloh}.
Letting $\sigma _{1}=I$, $\sigma _{2}=(1,5)$, and $\sigma _{3}=(1,3)$,
it can be shown that the above four states belong to the
families $\mathcal{F}_{2,2,2}^{\sigma _{1}\sigma _{2}\sigma _{3}}$, $%
\mathcal{F}_{3,3,3}^{\sigma _{1}\sigma _{2}\sigma _{3}}$, $\mathcal{F}%
_{2,4,2}^{\sigma _{1}\sigma _{2}\sigma _{3}}$ and $\mathcal{F}%
_{2,4,4}^{\sigma _{1}\sigma _{2}\sigma _{3}}$, respectively. Therefore,
these five-qubit states can also be distinguished by ranks.
Furthermore, it has been shown that five six-qubit states
$|\Xi _{2}\rangle$, $|\Xi _{4}\rangle$, $|\Xi _{5}\rangle$,
$|\Xi _{6}\rangle$, and $|\Xi _{7}\rangle$
are distinguished by the six-qubit filters \cite{Osterloh}.
Letting $\sigma_{1}=I$, $\sigma _{2}=(1,4)$, and $\sigma_{3}=(1,5)$,
it can be shown that the above five states belong to the families
$\mathcal{F}_{2,2,2}^{\sigma _{1}\sigma _{2}\sigma _{3}}$,
$\mathcal{F}_{2,2,4}^{\sigma_{1}\sigma _{2}\sigma _{3}}$,
$\mathcal{F}_{2,4,4}^{\sigma _{1}\sigma_{2}\sigma _{3}}$,
$\mathcal{F}_{3,4,4}^{\sigma _{1}\sigma _{2}\sigma _{3}}$,
and $\mathcal{F}_{3,3,3}^{\sigma _{1}\sigma _{2}\sigma _{3}}$, respectively.
Therefore, these six-qubit states can also be distinguished by ranks.

\section{V. Discussion and summary}

Chterental \emph{et al.} (see Remark 3.5 in \cite{Chterental}) stated that
the
family $L_{ab_{3}}$ is equivalent to a subfamily of $L_{abc_{2}}$ obtained
by setting $a=c$, where $L_{ab_{3}}$ and
$L_{abc_{2}}$ are given by \cite{Moor2}
\begin{eqnarray}
L_{ab_{3}} &=&a(|0000\rangle +|1111\rangle)
+\frac{a+b}{2}(|0101\rangle +|1010\rangle) \notag \\
&&+\frac{a-b}{2}(|0110\rangle +|1001\rangle)
+\frac{i}{\sqrt{2}}(|0001\rangle+|0010\rangle \notag \\
& & +|0111\rangle +|1011\rangle ),  \\
L_{abc_{2}} &=&\frac{a+b}{2}(|0000\rangle +|1111\rangle)
+\frac{a-b}{2}(|0011\rangle+|1100\rangle ) \notag \\
&&+c(|0101\rangle +|1010\rangle )+|0110\rangle .
\end{eqnarray}
In terms of the rank, the families $L_{ab_{3}}$ and $L_{abc_{2}}$ with $a=c$
are both divided into four subfamilies, see Table \ref{table_chterental}.
As can be seen, the subfamily $\mathcal{F}_{2}$ of $L_{ab_{3}}$ is a
single class with representative $\frac{i}{\sqrt{2}}(|0001\rangle
+|0010\rangle
+|0111\rangle +|1011\rangle )$, whereas the subfamily $\mathcal{F}_{2}$
of $L_{abc_{2}}(a=c)$ is a single class with representative
$\frac{b}{2}(|0000\rangle +|1111\rangle -|0011\rangle -|1100\rangle
)+|0110\rangle $. In light of Theorem 1 in \cite{LDF07a}, the two
representative states are not equivalent to each other. This reveals that $%
L_{ab_{3}}$ is not equivalent to a subfamily of $L_{abc_{2}}$ obtained by
setting $a=c$.

\begin{table}[tbph]
\caption{SLOCC classifications of $L_{ab_{3}}$ and $L_{abc_{2}}$}
\label{table_chterental}%
\begin{ruledtabular}
\begin{tabular}{ccccc}
$L_{ab3}$ & $\mathcal{F}_{1}$ & $\mathcal{F}_{2}$
& $\mathcal{F}_{3}$ & $\mathcal{F}_{4}$ \\
& $\emptyset $ & $a=b=0$
& $ab=0$ \& $a\neq b$ & $ab\neq 0$ \\
$L_{abc_{2}}$ & $\mathcal{F}_{1}$ &
$\mathcal{F}_{2}$ & $\mathcal{F}_{3}$ & $\mathcal{F}_{4}$ \\
$(a=c)$ & $a=b=0$ & $a=0$ \& $b\neq 0$
& $a\neq 0$ \& $b=0$ & $ab\neq 0$
\end{tabular}
\end{ruledtabular}
\end{table}

To determine if a four-qubit state belongs to a family
according to the criteria given by Verstraete \emph{et al.} \cite{Moor2}
and Lamata \emph{et al.} \cite{Lamata07},
one needs to check if the state is equivalent to the representative state of
that family.
For the classification scheme proposed in this Letter,
to determine if an $n$-qubit state belongs to a family,
one needs only to calculate the rank of the coefficient matrix of the state.

In summary, we have studied SLOCC classification for general $n$-qubit
states via the invariance of the rank of the coefficient matrix and given
several examples for $n$ up to six.
We have also characterized full separable states and genuinely entangled
states in terms of the rank.
We expect that the proposed entanglement classification for general $n$-qubit
states may find further experimental consequences.

This work was supported by NSFC (Grant No. 10875061) and
Tsinghua National Laboratory for Information Science and Technology.


\begin{thebibliography}{99}
\bibitem{Nielsen} M.A. Nielsen and I.L. Chuang, \textsl{Quantum Computation
and Quantum Information} (Cambridge Univ. Press, Cambridge, 2000).

\bibitem{Dur} W. D\"{u}r, G. Vidal, and J.I. Cirac, Phys. Rev. A 62, 062314
(2000).

\bibitem{Moor2} F. Verstraete, J. Dehaene, B. De Moor, and H. Verschelde,
Phys. Rev. A 65, 052112 (2002).

\bibitem{Chterental} O. Chterental and D.Z. Djokovi\'{c}, in Linear Algebra
Research Advances, edited by G.D. Ling (Nova Science Publishers, Inc.,
Hauppauge, NY, 2007), Chap. \textbf{4}, 133.

\bibitem{Lamata07} L. Lamata, J. Le\'{o}n, D. Salgado, and E. Solano, Phys.
Rev. A 75, 022318 (2007).

\bibitem{Borsten} L. Borsten, D. Dahanayake, M.J. Duff, A. Marrani, and W.
Rubens, Phys. Rev. Lett. 105, 100507 (2010).

\bibitem{Cao} Y. Cao and A.M. Wang, Eur. Phys. J. D 44, 159 (2007).

\bibitem{LDF07b} D. Li, X. Li, H. Huang, and X. Li, Phys. Rev. A 76, 052311
(2007);

\bibitem{LDFQIC} D. Li, X. Li, H. Huang, and X. Li, Quantum Inf. Comput. 9,
0778 (2009).

\bibitem{Buniy} R.V. Buniy and T.W. Kephart, arXiv:1012.2630.

\bibitem{Viehmann} O. Viehmann, C. Eltschka, and J. Siewert, Phys. Rev. A
83, 052330 (2011).

\bibitem{Zha} X. Zha and G. Ma, Chin. Phys. Lett. 28, 020301 (2011).

\bibitem{Chen} L. Chen and Y.X. Chen, Phys. Rev. A 74, 062310 (2006).

\bibitem{LDFEPL} D. Li, X. Li, H. Huang, and X. Li, Europhys. Lett. 87,
20006 (2009).

\bibitem{Bastin} T. Bastin, S. Krins, P. Mathonet, M. Godefroid, L. Lamata,
and E. Solano, Phys. Rev. Lett. 103, 070503 (2009).

\bibitem{Aulbach} M. Aulbach, arXiv:1103.0271.

\bibitem{LDFJPA} X. Li and D. Li, J. Phys. A: Math. Theor. 44, 155304 (2011).

\bibitem{LDFJPA12} X. Li and D. Li, J. Phys. A: Math. Theor. 45, 075308 (2012).

\bibitem{LDFQIC11} X. Li and D. Li, Quantum Inf. Comput. 11, 0695 (2011).

\bibitem{LDF07a} D. Li, X. Li, H. Huang, and X. Li, Phys. Rev. A 76, 032304
(2007).

\bibitem{Ribeiro} P. Ribeiro and R. Mosseri, Phys. Rev. Lett. 106, 180502
(2011).

\bibitem{Stockton} J.K. Stockton, J.M. Geremia, A.C. Doherty, and H.
Mabuchi, Phys. Rev. A 67, 022112 (2003).

\bibitem{Toth} G. T\'{o}th, J. Opt. Soc. Am. B 24, 275 (2007).

\bibitem{Huber} M. Huber, P. Erker, H. Schimpf, A. Gabriel, and B. Hiesmayr,
Phys. Rev. A 83, 040301(R) (2011).

\bibitem{Prevedel} R. Prevedel, G. Cronenberg, M.S. Tame, M. Paternostro, P.
Walther, M.S. Kim, and A. Zeilinger, Phys. Rev. Lett. 103, 020503 (2009).

\bibitem{Wieczorek} W. Wieczorek, R. Krischek, N. Kiesel, P. Michelberger,
G. T\'{o}th, and H. Weinfurter, Phys. Rev. Lett. 103, 020504 (2009).

\bibitem{Coecke} B. Coecke and A. Kissinger, arXiv:1002.2540.

\bibitem{Osterloh} A. Osterloh and J. Siewert, Int. J. Quant. Inf. 4, 531
(2006).
\end{thebibliography}
\end{document}